\documentclass[lettersize,journal]{IEEEtran}
\usepackage{amsmath,amsfonts}
\usepackage{array}
\usepackage{textcomp}
\usepackage{stfloats}
\usepackage{url}
\usepackage{verbatim}
\usepackage{graphicx}
\usepackage{multirow}
\usepackage{makecell}
\usepackage{float}
\usepackage{hyperref}
\usepackage{subfigure}
\usepackage{algorithmic}
\usepackage{algorithm}
\usepackage{booktabs}

\def\BibTeX{{\rm B\kern-.05em{\sc i\kern-.025em b}\kern-.08em
    T\kern-.1667em\lower.7ex\hbox{E}\kern-.125emX}}
\usepackage{balance}
\begin{document}
\title{LiPar: A Lightweight Parallel Learning Model for Practical In-Vehicle Network Intrusion Detection}
\author{Aiheng Zhang,~Qiguang Jiang,~Kai~Wang$^*$, Ming~Li

\thanks{
This work was supported in part by National Natural Science Foundation of China (NSFC) (grant number 62272129) and Taishan Scholar Foundation of Shandong Province (grant number tsqn202408112).
\textit{(Corresponding author: Kai Wang.)}}
\thanks{Aiheng Zhang, Qiguang Jiang, and Kai Wang are with the School of Computer Science and Technology, Harbin Institute of Technology, Weihai, China. (e-mail: zahboyos@163.com; jiangqiguang\_971@163.com; dr.wangkai@hit.edu.cn)

Ming~Li is with Shandong Inspur Database Technology Co., Ltd, China. (email: liming2017@inspur.com) }
}
\markboth{}%
{Shell \MakeLowercase{\textit{et al.}}: Aiheng Zhang}

\maketitle

\begin{abstract}
With the development of intelligent transportation systems, vehicles are exposed to a complex network environment. As the main network of in-vehicle networks, the controller area network (CAN) has many potential security hazards, resulting in higher generalization capability and lighter security requirements for intrusion detection systems to ensure safety. Among intrusion detection technologies, methods based on deep learning work best without prior expert knowledge. However, they all have a large model size and usually rely on large computing power such as cloud computing, and are therefore not suitable to be installed on the in-vehicle network. Therefore, we explore computational resource allocation schemes in in-vehicle network and propose a lightweight parallel neural network structure, LiPar, which achieve enhanced generalization capability for identifying normal and abnormal patterns of in-vehicle communication flows to achieve effective intrusion detection while improving the utilization of limited computing resources. In particular, LiPar adaptationally allocates task loads to in-vehicle computing devices, such as multiple electronic control units, domain controllers, computing gateways through evaluates whether a computing device is suitable to undertake the branch computing tasks according to its real-time resource occupancy. Through experiments, we prove that LiPar has great detection performance, running efficiency, and lightweight model size, which can be well adapted to the in-vehicle environment practically and protect the in-vehicle CAN bus security. Furthermore, with only the common multi-dimensional branch convolution networks for detection, LiPar can have a high potential for generalization in spatial and temporal feature fusion learning.
\end{abstract}


\section{Introduction}
\label{Section1}
\IEEEPARstart{W}{ith} the development of mobile communication technology and Internet of Things (IoT) technology, the Internet of vehicles and in-vehicle networks (IVNs) have been widely applied in key functions of intelligent transportation systems. 
The Internet of vehicles can help vehicles acquire environmental information promptly and make vehicles more intelligent by sharing information with neighboring vehicles and infrastructure such as roadside surveillance \cite{Comprehensive-Survey-for-safe-efficient}.  However, increased vehicular connectivity with external terminals provides more opportunity for external attackers and results in more risk of being attacked \cite{Potential-cyberattacks,attack-surfaces}.

In-vehicle communication among Electronic Control Units (ECUs) is mainly transmitted via the Controller Area Network (CAN), which is a standard backbone network of in-vehicle communication. CAN is a message-broadcast-based protocol without node information for the sender and receiver. It lacks authentication information and message encryption mechanisms, making it easy for attackers to intrude and corrupt CAN communication \cite{Survey-Automotive-CAN-IDS,cyberattacks-countermeasures}. Therefore, improving the security of the in-vehicle CAN bus is an important aspect to ensure vehicle security and passenger safety.

The intrusion detection system (IDS) has always been a focus of research in the field of vehicle security. IDS can detect attacks timely and accurately, so as to help vehicle-mounted systems make appropriate responses in time. As a result, intrusion detection is of paramount importance for in-vehicle CAN bus \cite{Survey-CAN-CC}. Current intrusion detection techniques can mainly fall into four categories \cite{TVT-VehicularIDS-PseudoData}: (1) the specification-based approaches, (2) the fingerprint-based approaches, (3) the statistics-based approaches, and (4) the learning-based approaches. Among them, the dominant techniques are learning-based approaches.
Deep learning-based approaches, in particular, have better real-time detection performance in the face of continuously changing attacks through multiple training. 
The reasons are that deep learning-based approaches can provide more intelligent IDS to detect attacks without prior expert knowledge \cite{CFP-VehicularSecurity-DL, VehCom-AI-VANET}. Compared with the other approaches, deep learning-based methods have the capability to handle huge amounts of data, learn attack characteristics, predict the occurrence of attacks, and achieve more accurate detection results \cite{Survey-IDS-IVN}.

In recent years, most of the in-vehicle CAN bus intrusion detection technologies are developed based on Convolutional Neural Networks (CNN) and Long Short-Term Memory (LSTM) networks \cite{VehCom-VehicularIDS-DCNN,SAC-VehicularIDS-CANTransfer}, due to the multidimensional time series CAN data. 
Although these methods have extremely high detection accuracy, they all use linear stacking of neural network layers to build models, which are usually characterized by the large model size. They consume considerable computing resources and may lead to a large delay in detection. 

Although deep learning models generally rely on the Cloud Computing (CC) server or Mobile Edge Computing (MEC) server which has sufficient resources, they are not as reliable and timely as local computing \cite{local-cloud-computing}. In fact, vehicles have extremely high requirements for the reliability and timeliness of intrusion detection technology, and the computing resources in the in-vehicle environment are very limited \cite{IVN-computing-resource}. A linear stacked deep learning model might only run on one ECU device or one in-vehicle computing component, which may cause overload and long delay, and affect the original functions of the vehicle. In our previous study \cite{our-analysis}, we selected 10 representative intrusion detection models and evaluated their adaptability for the in-vehicle environment, and proposed corresponding baseline selection schemes, which provided a basis for this study.

To solve the above problems, we propose a novel lightweight method for IVN intrusion detection. The main contributions of our work can be summarized as follows:
\begin{itemize}
	\item We design a lightweight parallel network model (LiPar), which is the first parallel-based intrusion detection method for IVN. The model allocates the task load to multiple parallel branches and multiple computing devices according to a resource adaptation algorithm, and adopts lightweight structures to compose each branch.
	\item LiPar is able to identify normal and abnormal patterns embedded in time series data of in-vehicle communication flows accurately. Specifically, a multidimensional feature extraction approach based on the combination of CNN and LSTM structures is designed to make information fusion from spatial and temporal views to improve the generalization capability.
	\item A resource adaptation algorithm is designed to make full use of multi computing resources, which can evaluate whether a computing device is suitable to undertake the branch computing tasks according to its real-time resource occupancy.
	\item Extensive experimental evaluation and comparative analysis are conducted to demonstrate that the proposed LiPar has outstanding detection performance, running efficiency, and a lightweight degree, which is significant to be applied in IVN.  
\end{itemize}

The paper is organized as follows. Section~\ref{Section2} introduces the important background knowledge related to our research. Section~\ref{Section3} explains the proposed framework and design details. Section~\ref{Section4} describes the experimental settings, results, and evaluation. Section~\ref{Section5} summarizes the related research work in this field, and Section~\ref{Section6} concludes this paper.


\section{Background}\label{Section2}
\subsection{In-vehicle controller area network}
The in-vehicle network is important to realize complex functions, high performance, and good comfort of vehicles. The structure of the in-vehicle network is related to the manufacturer and the type of vehicle and has no unified standard \cite{Chinese-Conference}, most of which is mainly implemented by CAN, LIN, and Ethernet. According to the development trend of in-vehicle networks, CAN is in the dominant position.

\begin{figure*}[tbp]
\centering
\includegraphics[width=5.3in]{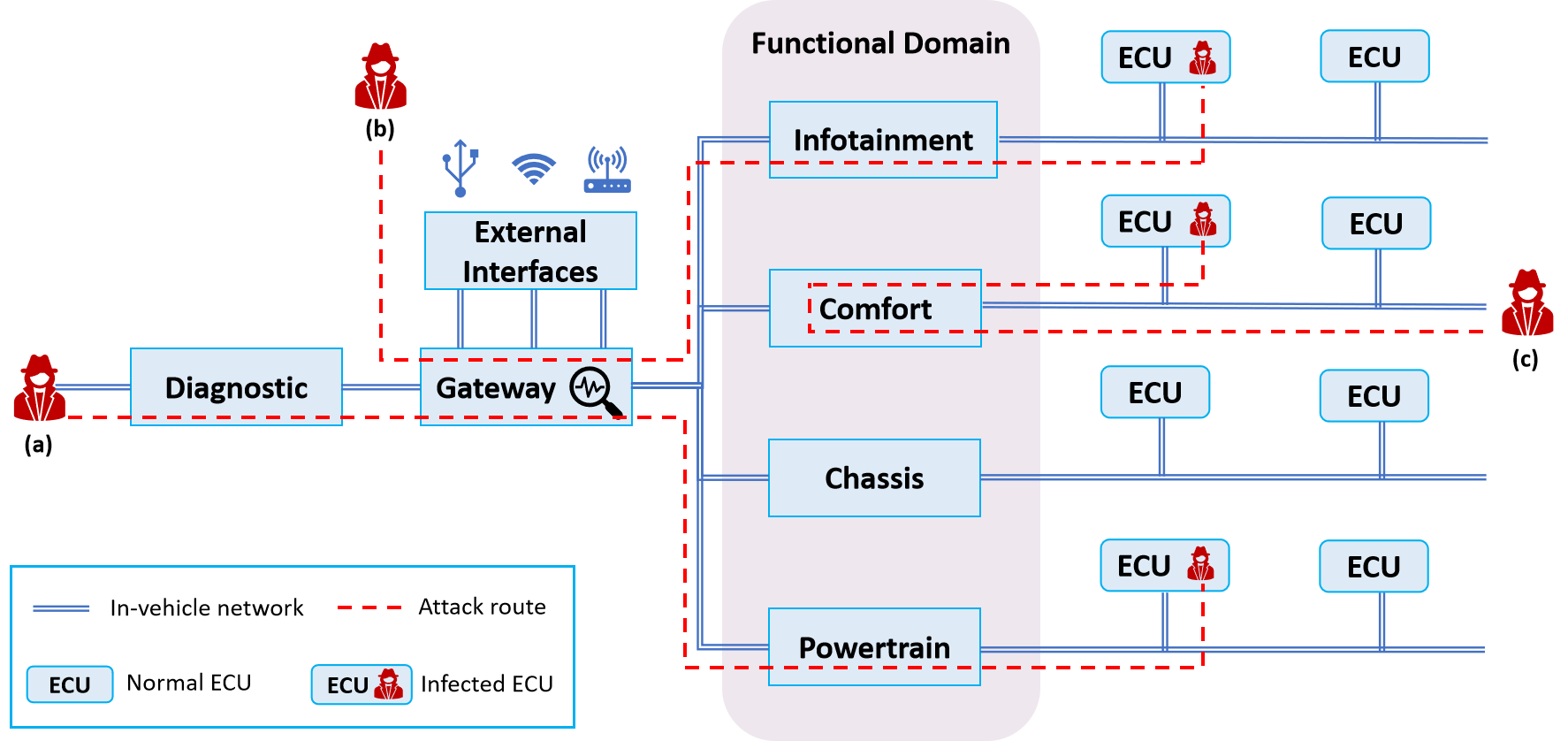} 
\caption{The topology of in-vehicle CAN bus and the possible attack routes: (a) attacking through the OBD-II port; (b) attacking from the external interfaces; (c) attacking by infected ECU to occupy the CAN bus.}
\label{fig_topology}
\end{figure*}

CAN is developed by Robert Bosch in 1985 to reduce communication costs and complexity. Due to its high efficiency and stability, CAN has been the most representative in-vehicle network technology and is widely used in the On-Board Diagnostics II (OBD-II) standard as a major protocol \cite{sensor-hierarchical}. As shown in Fig.~\ref{fig_topology}, the topology of in-vehicle CAN can be divided into different control units according to their functional domains, such as the infotainment units, comfort units, chassis units, and powertrain units. Benefiting from the message-broadcast-based characteristic of the in-vehicle CAN bus, the ECUs can be interconnected with each other and make efficient cooperation. However, there is only one gateway between the external interfaces, diagnostic, and the CAN bus which has low security without permission authentication and information encryption. Fig.~\ref{fig_topology} also illustrates that the OBD-II port and the external interfaces, such as Bluetooth, WiFi, and telematic services, have provided a wide attack surface to the in-vehicle CAN bus. Once the attacker breaks through the gateway, the entire in-vehicle network will be threatened. Therefore, the CAN bus is very fragile and is surrounded by various threats, and the gateway is the most fatal place to install the IDS.

\begin{figure*}[tbp]
\centering
\includegraphics[width=5.2in]{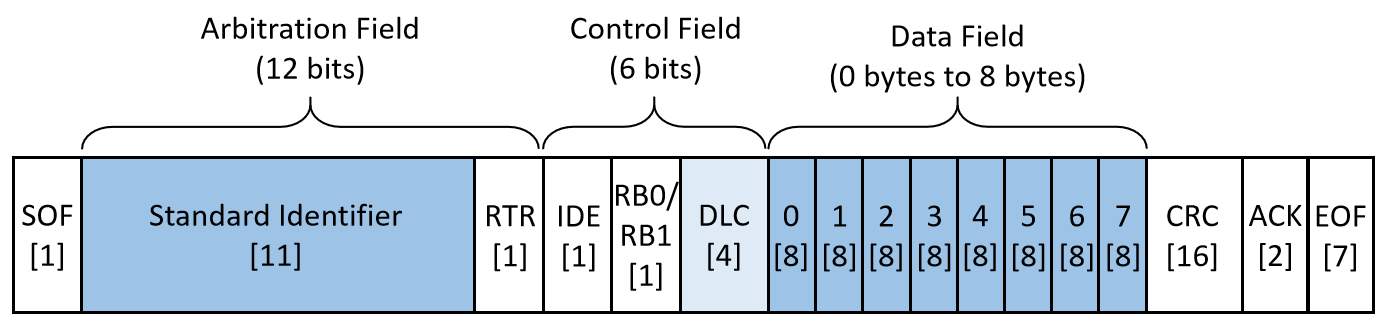} 
\caption{The standard data frame of CAN bus.}
\label{fig_data_frame}
\end{figure*}

Generally, the in-vehicle CAN protocol has two formats: the standard format and the extended format. The standard format was standardized by the International Organization for Standardization (ISO) 11898 \cite{ISO}, as shown in Fig.~\ref{fig_data_frame}, which has an 11-bit identifier (ID), while the extended format has a 29-bit ID frame. The standard data frame is composed of a 1-bit Start of Frame (SOF), a 12-bit arbitration field which consists of an 11-bit ID and 1-bit RTR, a 6-bit control field mainly including a 4-bit DLC, a data field (called payload) in the range of 0 to 8 bytes (the length is related to the value of DLC), a CRC field, an ACK field, and a 7-bit End of Frame. 

In the field of intrusion detection, researchers usually pay more attention to the CAN ID, DLC, and payload. They are the three most important data fields for vehicle control and sensor feedback and will contain more attack features if an attack occurs. The CAN ID determines the priority of CAN messages, the DLC determines the length of the data field and the payload contains most of the control information and status information of ECUs. A representative and widely used dataset for in-vehicle CAN intrusion detection, called the Car-Hacking dataset, is collected by Seo \textit{et al} \cite{GIDS}. Some samples in the Car-Hacking dataset with different values of ID, DLC, and data payload are shown in Table~\ref{table_sample_data}.

\begin{table}[tbp]
\renewcommand{\arraystretch}{1.3}
\caption{Sample of CAN packets: different ID, DLC, and payload fields}
\label{table_sample_data}
\centering
\resizebox{0.48\textwidth}{!}
{
\begin{tabular}{llll}
\toprule
\textbf{Timestamp} & \textbf{ID} & \textbf{DLC} & \textbf{Payload}\\
\midrule
1479121434.854108 & 0545 & 8 & d8 00 00 8a 00 00 00 00\\
1479121434.854290 & 02b0 & 5 & 8d ff 00 07 02\\
1479121434.854947 & 043f & 8 & 00 40 60 ff 5a 6c 08 00\\
1479121434.869396 & 05f0 & 2 & f4 00\\
1479121434.870212 & 0350 & 8 & 05 28 a4 66 6d 00 00 82\\
\bottomrule
\end{tabular}
}
\end{table}

In addition to the normal in-vehicle CAN communication message data, there are four types of network-based attacks in this dataset, DoS attack, fuzzy attack, and impersonation attack, including spoofing the drive gear and spoofing the RPM gauze. They can all attack vehicles via network connections and have different attacking behaviors. The details of these attacks are given as follows.

\begin{itemize}
	\item [1)]
	\emph{DoS attack}: An attack that blocks other communication information by occupying the communication channel. Since the CAN bus structure is based on message broadcasting, the order of message sending is determined by the priority of messages. The lower the value of the CAN ID, the higher the priority of the CAN message. Therefore, a DoS attacker always forges numerous CAN messages with the lowest ID values of 0x0000 to occupy the CAN bus resources and cause system paralysis.
	\item [2)]
	\emph{Fuzzing attack}: An attack that uses malicious ECU to send fake messages to the in-vehicle CAN bus. Fuzzy attacks use random ID values to reduce the recognition of attack features and avoid being detected. Unlike DoS, fuzzy attacks are much slower to send fake messages. If the sending speed is the same as the normal CAN messages, this type of attack will be difficult to be discovered.
	\item [3)]
	\emph{Impersonation attack}: An attack that can implement unauthorized access by spoofing legitimate authentication. The attacker will simulate a legitimate ECU node and send effective simulated messages to the in-vehicle CAN bus, but cause the component behavior to be abnormal. Spoofing the drive gear and spoofing the RPM gauze are two attacks to simulate the drive gear and RPM behavior, respectively.
\end{itemize}

In order to improve the security of in-vehicle networks, this paper proposed a novel intrusion detection technology that is lightweight enough to be installed on the gateway. And, its detection effect is evaluated on the four types of attack data in the Car-Hacking dataset mentioned above. The details of experimental settings and results are described in Section~\ref{Section4}.

\subsection{Electronic control units}
The in-vehicle CAN is actually designed to provide more efficient and faster communication between ECUs. Actually, ECU is indispensable to the realization of various complex functions in modern intelligent vehicles. 

\begin{figure}[tbp]
\centering
\includegraphics[width=3in]{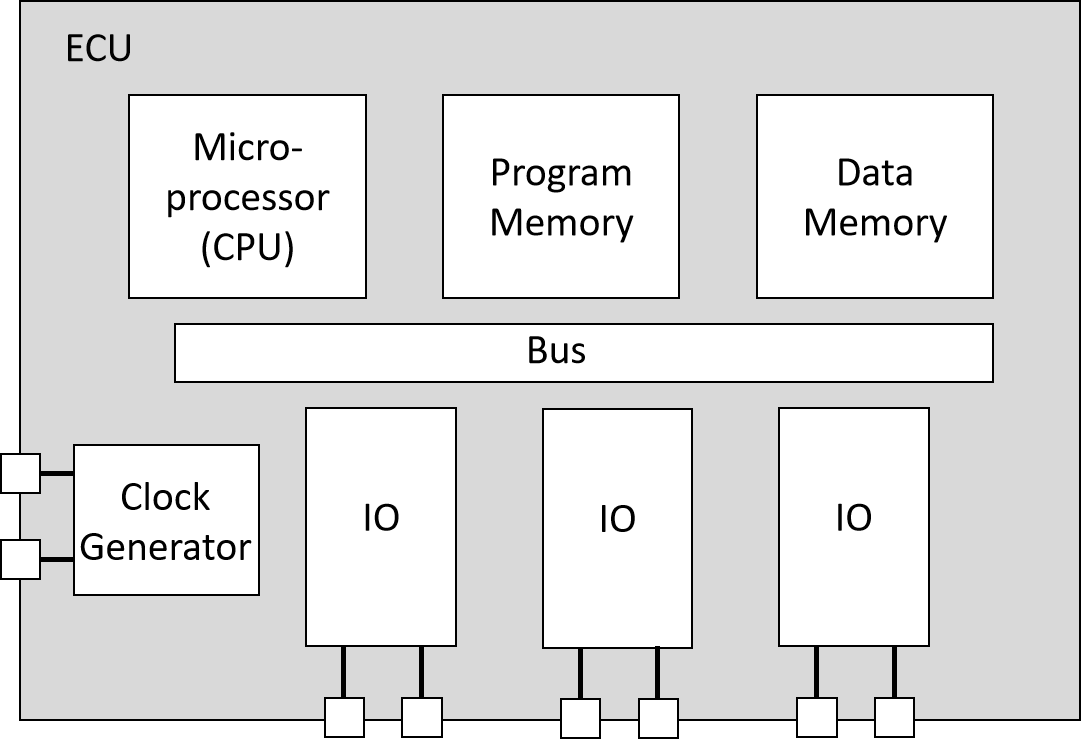} 
\caption{The schematic diagram of ECU internal structure.}
\label{fig_ECU}
\end{figure}

There are many sensors and actuators in the vehicle, which are respectively used to obtain the status data of vehicle components and make the components perform the specified actions according to the commands. The ECU is responsible for receiving the electrical signals from the sensor, calculating and evaluating these electrical signals according to the control program set in advance. In addition, if the ECU receives a control command from the driver, it will also calculate the triggering signal for the actuator according to the program. As shown in Fig.~\ref{fig_ECU}, the control program is stored in a special memory, and the calculation is done by a microprocessor in the ECU. Besides, it mainly contains some data memory, IO ports, and a clock generator. On the whole, the internal structure of ECU is very simple, but this also leads to very limited memory and computing resources \cite{ECU}, especially for running and storing a deep learning model which has a large size and computation.

In the automotive industry, companies and manufacturers are equipping their vehicles with a large number of ECUs to control various vehicle components, such as ventilation systems, window control systems, and engine control systems, in order to support advanced system functions \cite{Survey-CAN-CC}. Generally, the number of ECUs is about dozens in the vehicle and can be as high as hundreds in a high-end automobile. Also, these ECUs are serial on their own in-vehicle buses as illustrated in Fig.~\ref{fig_topology}. Therefore, in order to solve the resource limitation problem of a single ECU, we designed a parallel lightweight neural network based on the above characteristics to make full use of multiple ECU resources.

In fact, different ECU resources in the car have different utilization and importance. For example, as shown in Fig.~\ref{fig_topology}, the ECUs can be divided into different domains according to the functional modules they are responsible for. When considering vehicle security, the ECUs in the powertrain domain must be more important than the ECUs in the infotainment domain. Considering the frequency of utilization, the ECUs in the chassis domain, like the brake control unit, may be busier than ECUs in the comfort domain, like the window control unit. Because different ECUs have different risk indexes and resource occupancy, we also designed an algorithm to dynamically allocate the running load of the model according to the ECU status.

\subsection{Convolutional neural network}
The deep neural network uses high-level features to represent the abstract semantics of data by building a multi-layer network, so it has excellent feature learning ability, but also has a large memory consumption and computing resource consumption. CNN is a kind of deep neural network with convolution structures, but convolution structures can reduce the amount of memory occupied by a neural network. There are three key operations: receptive field localization, weight sharing, and pooling, which effectively reduce the number of network parameters and alleviate the problem of overfitting the model. As illustrated in Fig.~\ref{fig_conv}:

\begin{figure}[tbp]
\centering
\includegraphics[width=3in]{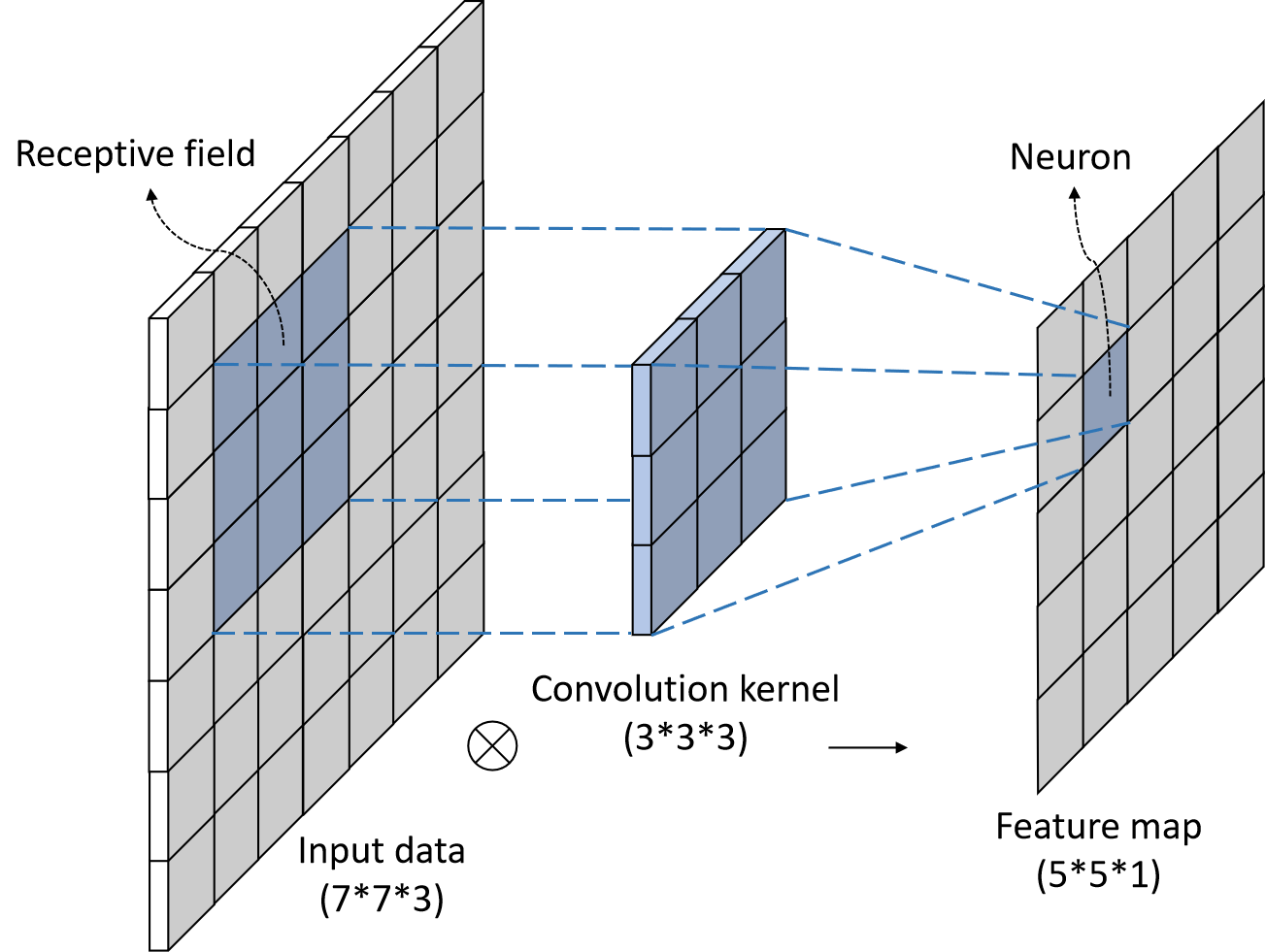} 
\caption{The schematic diagram of convolution calculation. The convolution kernel is used as the weight to multiply and add the corresponding input data pixel points to obtain a neuron of the feature map. Then, the convolution kernel is sliding according to the step size (1 in this figure) to calculate other neurons, forming a complete feature map.}
\label{fig_conv}
\end{figure}

\begin{itemize}
	\item Receptive field localization: each neuron does not need to receive the whole image, but only needs to feel the local features. Then, the global information can be obtained by synthesizing these different local neurons at a higher level, which can reduce the number of connections.
	\item Weight sharing: A convolution operation of an image uses the same convolution kernel for each neuron. That can reduce the parameters to be calculated, and different feature maps (also called channels) can be obtained by convoluting the image with multiple convolution kernels. Its main ability is to detect the same type of features at different positions, that is, it has good translation invariance.
	\item Pooling: It mainly includes two operations, average pooling and maximum pooling, which take the average or maximum value of all pixels in the receptive field as a neuron of the feature map. Its main function is to select features and reduce the number of features, thus reducing the number of parameters.
\end{itemize}

Furthermore, we can get the formula for calculating the size of the receptive field from the convolution operations:
\begin{equation}
F\left(i \right) = \left ( F\left (i+1  \right ) -1\right ) \times Stride + Ksize,
\label{eq_conv1} 
\end{equation}
where $F\left (i  \right )$ is the side length of the receptive field on the $i^{th}$ level, $Stride$ is the step size of sliding on the $i^{th}$ level and 
$Ksize$ is the side length of the convolution or pooling kernel. According to the Equation (\ref{eq_conv1}), some research has proved that the $5 * 5$ convolution kernel can be replaced by stacking two $3 * 3$ convolution kernels, and the $7 * 7$ convolution kernel can be replaced by stacking three $3 * 3$ convolution kernels with the same size of the receptive field \cite{VGG-kernelsize}. Obviously, the $3*3$ kernel has much fewer parameters than a larger-scale kernel. Therefore, in our model design, we stack multiple layers of $3*3$ convolution kernels instead of a large-scale convolution kernel.

\subsection{Recurrent neural network}
\begin{figure}[tbp]
\centering
\includegraphics[width=2.5in]{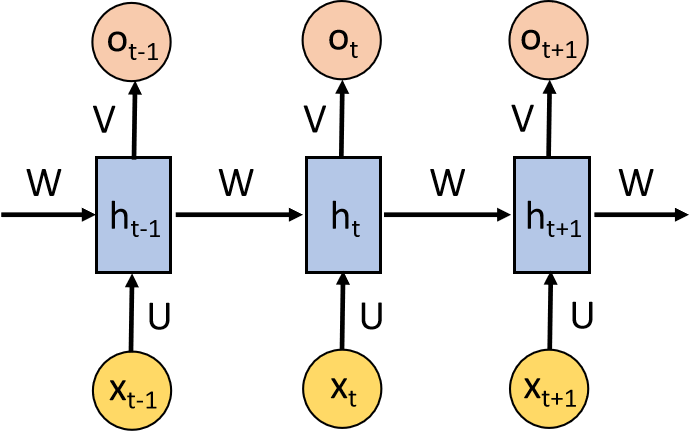} 
\caption{The schematic diagram of recurrent neural networks.}
\label{fig_rnn}
\end{figure}

The convolutional neural network can effectively learn the spatial features of each input, but it can not learn the temporal features, which is the sequence relationship between inputs. The Recurrent Neural Network (RNN) is a kind of neural network with memory ability \cite{rnn}. In the RNN, neurons can not only learn their own input features but also receive information from other neurons, forming a long-term ``memory''. As shown in Fig.~\ref{fig_rnn}, $x_{t}$ is a vector, which represents the value of the input layer at time $t$. $h_{t}$ is a vector representing the value of the hidden layer at time $t$. $o_{t}$ is a vector representing the value of the output layer at time $t$. $U$ and $V$ are the weight matrices respectively for the computation between the input layer and the hidden layer, and between the hidden layer and the output layer. The weight matrix $W$ is for the transfer of information contained in the hidden layer between different times. As a result, the value $h$ of the hidden layer depends not only on the current input $x$ but also on the value $h$ at the previous time, which contains all the information of $h$ at previous times. In the same hidden layer, the weight matrices $W$, $V$, and $U$ at different times are equal, which forms the parameter sharing in the RNN.

\section{The proposed framework}
\label{Section3}
\subsection{The problem statement}
Deep learning algorithms are widely used in many fields, but most of them are only available in cloud computing because of their large calculation and ample cloud resources. However, the computing resources on devices, especially ECU devices, are too limited to execute deep learning algorithms with large model sizes. The existing in-vehicle CAN bus intrusion detection technologies have the problems of massive model parameters, large size, high detection delay, and high resource consumption for one computing device. 

Actually, in the field of deep learning, there is always a contradiction between the lightweight degree and detection effect of a learning model, which is often impossible to have both. There is also a certain contradiction between the timeliness and complexity of the model and the detection effect of the model in the field of deep learning, as well as the field of intrusion detection. Also, high model complexity is always required to achieve a high detection accuracy but can lead to low timeliness. However, in-vehicle IDS has a high requirement for timeliness because even a very short breakdown can be fatal to a speeding car. Therefore, our goal is to design a learning model that can detect intrusion attacks with high accuracy, low time delay, and low resource consumption for each computing device.

In our study, in-vehicle CAN bus data is obtained in chronological order: the training data is denoted as $x_{train} = \left [ x_{train}^{(1)}, \dots, x_{train}^{(T)} \right ]$, which is used to train our models. The value $x_{train}^{t}$ is an $N$ dimensional vector representing the values of the $t^{th}$ in-vehicle CAN message with $N$ data fields. The training data consists of normal data and four types of attack data.
Our purpose is to detect attacks in testing data which is denoted as $x_{test} = \left [ x_{test}^{(1)}, \dots, x_{test}^{(T')} \right ]$. The output of our model is the corresponding set of multiple classification result labels.

\subsection{The framework overview}
According to the characteristics of the in-vehicle CAN bus, which is based on message broadcast and equipped with hundreds of ECUs and other computing devices. We make full use of its natural advantages of parallel resources and propose an intrusion detection method for the in-vehicle CAN bus based on parallel lightweight multi-feature learning architecture, which can dynamically adapt the running load to resources of multiple computing devices. Specifically, we present ECU devices as the example of the use for LiPar to allocate computational resources. In fact, LiPar is applicable to all scenarios where computing resources are allocated to multiple devices.

\begin{figure}[tbp]
\centering
\includegraphics[width=3.4in]{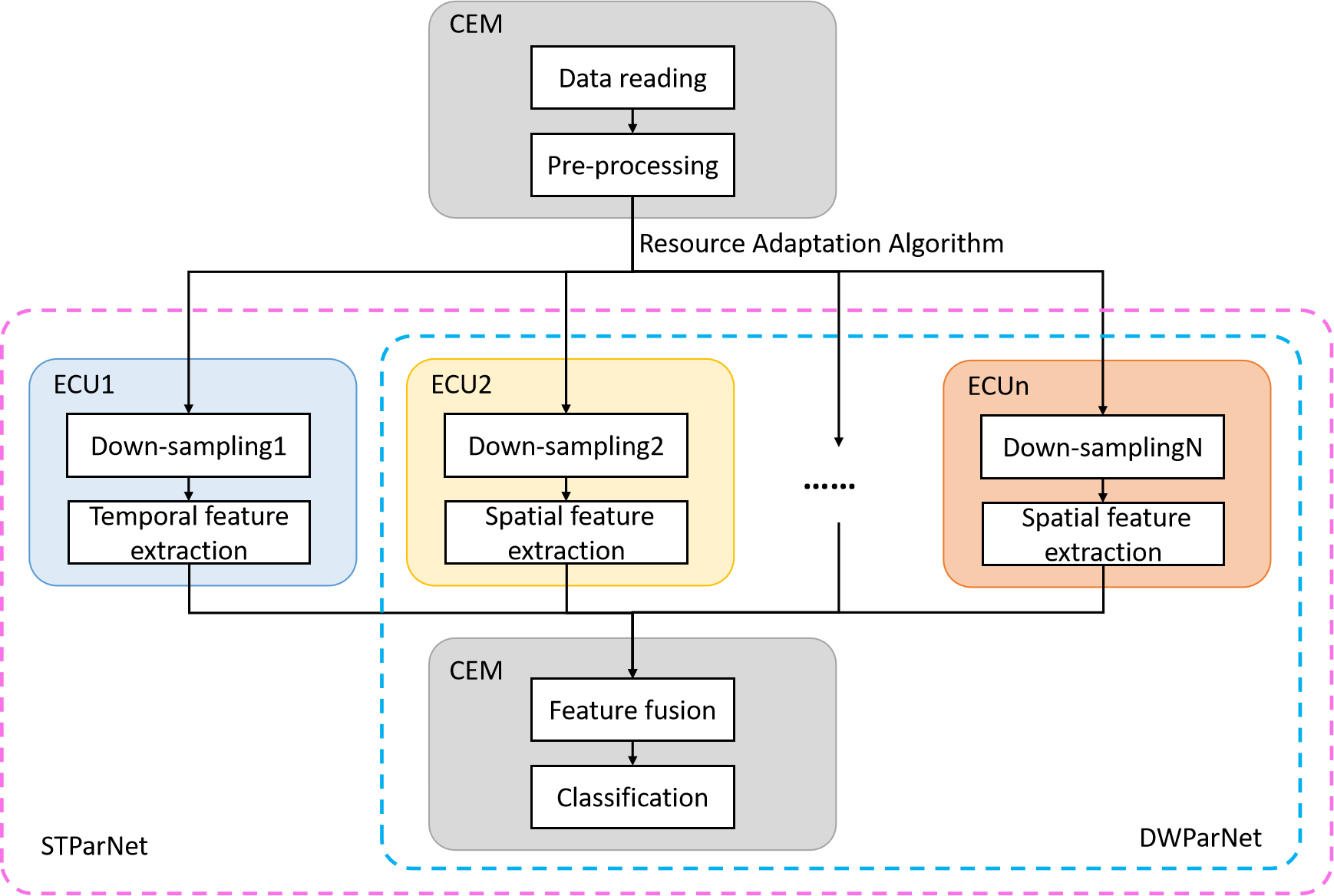} 
\caption{The architecture of our parallel lightweight learning method based on resource adaptation.}
\label{fig_LPNarct}
\end{figure}

Fig.~\ref{fig_LPNarct} shows the proposed parallel lightweight learning architecture, named LiPar. The steps of data reading, data pre-processing, and final feature fusion and classification of in-vehicle CAN messages are all completed in the central electronic module (CEM), which can be the OBD-II or the gateway of the in-vehicle CAN bus. Each branch network is the main network structure for feature extraction and feature learning, which contains a large amount of computation. Therefore, the computation tasks of each branch will be assigned to different ECUs according to the resource adaptation algorithm proposed in this study. Also, the network design of each branch is shallow (no more than 10 layers), so as to ensure that each branch network is lightweight enough and will not generate excessive load on ECU. In addition, the parallel structure of DWParNet can enable each branch to conduct down-sampling and spatial feature extraction from different dimensions. STParNet can use one more branch for temporal feature extraction to learn more comprehensive and full features and improve the detection performance of the model after feature fusion. 

\subsection{DWParNet: the parallel learning network for spatial feature extraction}
The DWParNet we proposed plays an important role in the whole model, mainly reflected in two aspects: 1) It undertakes almost all the tasks of extracting and learning spatial feature information, and 2) the main parts of the parallel structure are embodied in the DWParNet.

We adopt CNN structure in DWParNet for spatial feature extraction and limit the size of convolution kernels to $3 * 3$ instead of larger-scale kernels to reduce the number of training parameters. However, traditional convolution still has some redundant computations that can be pruned. For further lightweight design in DWParNet, we adopt a lot of Depthwise (DW) Convolutions, which is a special form of group convolution, instead of traditional convolution to reduce the model parameters and computation. 

\begin{figure}[tbp]
\centering
\subfigure[]{
\label{normal-conv} 
\includegraphics[width=0.3\linewidth]{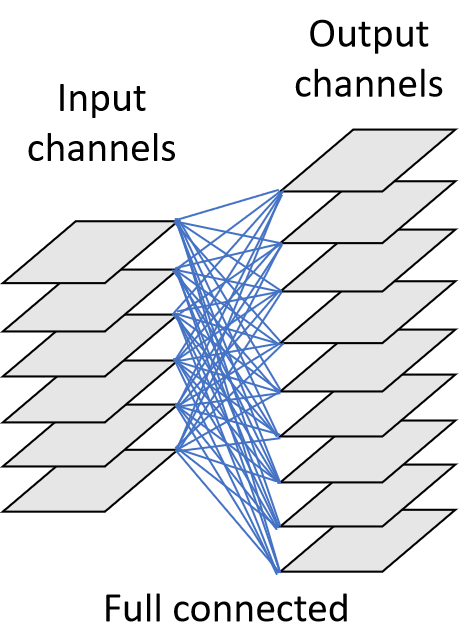}}
\subfigure[]{
\label{group-conv} 
\includegraphics[width=0.3\linewidth]{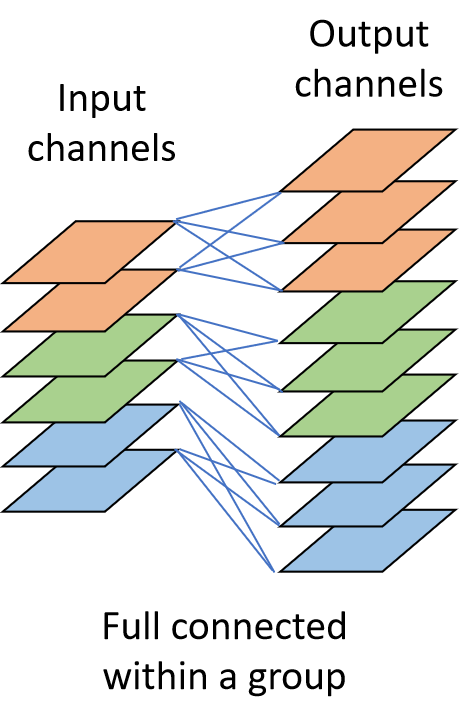}}
\subfigure[]{
\label{DW-conv} 
\includegraphics[width=0.3\linewidth]{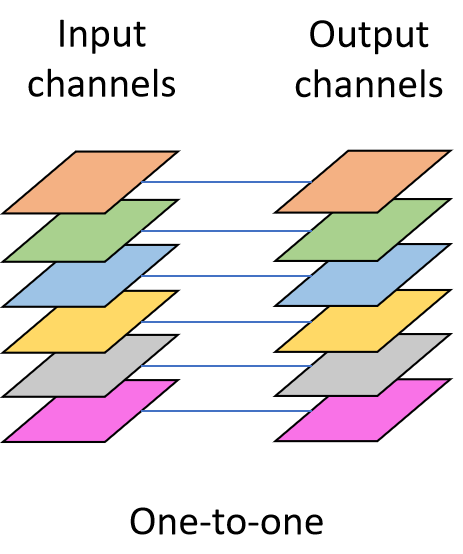}}
\caption{The schematic diagram of conventional convolution, grouping convolution, and DW convolution. (a) Each input channel is fully connected with $C_{o}$ kernels to get $C_{o}$ output channels. (b) Each color represents each group, in which $C_{i}/g$ input channels are fully connected with $C_{o}/g$ kernels to get $C_{o}/g$ output channels. (c) Each input channel is connected with one kernel to get one output channel.}
\label{fig_conv-type} 
\centering
\end{figure}

In general, the convolution on one layer will use multiple convolution kernels to obtain more features, the number of which is consistent with the number of convolution kernels. As Fig.~\ref{normal-conv} illustrated, each channel of the input image is fully connected with each convolution kernel. Furthermore, group convolution is proposed to reduce computation and parameters \cite{group-conv}, which divides the channels of the input data and convolution kernels into different groups for calculation, as shown in Fig.~\ref{group-conv}. For example, we assume that the input data shape is $H_{i} * W_{i} * C_{i}$; the output data shape is $H_{o} * W_{o} * C_{o}$; and the number of groups as $g$. Since the number of channels of convolution kernels is equal to the input channels and the number of convolution kernels is equal to the output features, the number of standard convolution parameters is 
\begin{equation}
N = Ksize * Ksize * C_{i} * C_{o},
\label{eq_conv2} 
\end{equation}
and the number of group convolution parameters is
\begin{equation}
Ksize * Ksize * \frac{C_{i}}{g} * \frac{C_{o}}{g} * g = N / g.
\label{eq_conv3} 
\end{equation}
Therefore, group convolution can greatly reduce the parameter amount according to the number of groups. Especially, when $C_{o} = C_{i}$, we can set $g = C_{i}$ and it becomes DW convolution as shown in Fig.~\ref{DW-conv}. The disadvantage of DW convolution is that it can only get the output features with the same shape as the input data.

\begin{figure*}[tbp]
\centering
\includegraphics[width=6.5in]{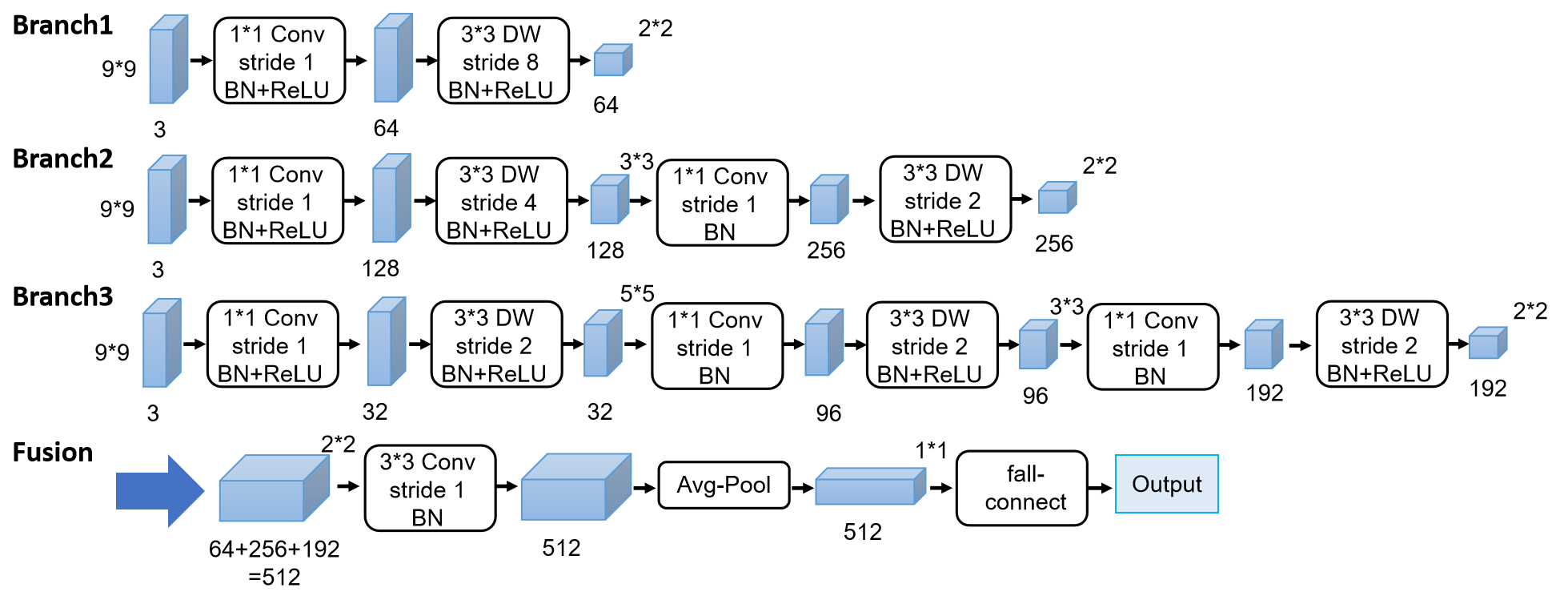} 
\caption{The network structure of DWParNet.}
\label{fig_dw}
\end{figure*}

To solve the drawbacks of DW convolution, we only use the traditional convolution operation with a convolution core size of $1 * 1$ in DWParNet when the number of channels needs to be adjusted by down-sampling. According to experimental experience, the number of network branches is set to 3. The details of DWParNet are shown in Fig.~\ref{fig_dw}. We use batch normalization (BN) to accelerate network training by reducing internal covariate and the ReLU activation function to prevent gradient vanishing during model training. Each branch adopts different granularity feature extraction from multi-dimension, but they all obtain $2 * 2$ multi-channel feature matrices so that they can be merged in the direction of the channel and learn more comprehensive features.

\subsection{STParNet: integrating temporal feature extraction}
In order to improve the detection effect and generalization ability of the model, we integrate the temporal feature learning structure on the basis of DWParNet. 

Although RNN is a classical and efficient temporal feature extraction network, during the training of the RNN, with the extension of training time and the increased number of network layers, the problem of gradient explosion or gradient disappearance occurs easily. To solve this problem, we adopt the LSTM network which is an improved scheme of RNN \cite{LSTM-proposed}. The LSTM network has improved the calculation between the input layer and the hidden layer by adding a memory cell $C_{t}$ to selectively memorize information. In addition, the LSTM network also sets a forgetting gate $f_{t}$, update gate $i_{t}$, and output gate $o_{t}$, to discard useless memory, learn the essence from the new input, memorize it, and send it to the next moment state $h_{t}$. The specific calculation process is as follows:
\begin{equation}
f_{t} = \sigma \left (W_{f}\cdot \left [h_{t-1}, x_{t}  \right ] + b_{f} \right ),
\label{eq_lstm1} 
\end{equation}
\begin{equation}
i_{t} = \sigma \left (W_{i}\cdot \left [h_{t-1}, x_{t}  \right ] + b_{i} \right ),
\label{eq_lstm2} 
\end{equation}
\begin{equation}
o_{t} = \sigma \left (W_{o}\cdot \left [h_{t-1}, x_{t}  \right ] + b_{o} \right ),
\label{eq_lstm3} 
\end{equation}
\begin{equation}
\tilde{C}_{t} = tanh \left (W_{C}\cdot \left [h_{t-1}, x_{t}  \right ] + b_{C} \right ) ,
\label{eq_lstm4}
\end{equation}
\begin{equation}
C_{t} = f_{t} * C_{t-1} + i_{t} * \tilde{C}_{t},
\label{eq_lstm5}
\end{equation}
\begin{equation}
h_{t} = o_{t} * tanh \left (C_{t} \right ),
\label{eq_lstm6} 
\end{equation}
where $W$ is the weight matrix of each gate, $b$ is the bias of each gate, and $\tilde{C}_{t}$ is the cellular state of the current memory cell $C_{t}$.

\begin{figure}[tbp]
\centering
\includegraphics[width=2in]{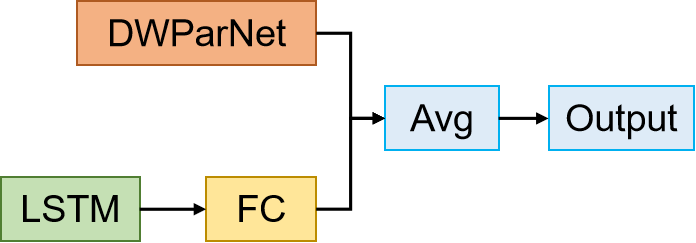} 
\caption{The schematic diagram of STParNet structure.}
\label{fig_st}
\end{figure}

To explore the temporal relationship between in-vehicle CAN messages, we add the LSTM network structure to the model, analyze it in combination with the spatial features extracted by DWParNet, and obtain the final model, STParNet, as shown in Fig.~\ref{fig_st}. The LSTM layer is mainly responsible for learning the temporal features of in-vehicle CAN messages. The full connection layer is to decode the hidden state of the last time step and get a logit, which is the output of the temporal feature learning branch. Then, we calculate the average of the logit obtained by DWParNet and the logit of temporal features and take it as the final classification basis.

After parameter adjustments and model optimization, we determine the hyperparameters that make the model performance optimal, as shown in Table~\ref{table_hyper_model}. Because the number of convolution cores in DW convolution operation is determined by the number of channels of input data, no human intervention is required, so it is not listed in Table~\ref{table_hyper_model}. Except for the convolution layer of the fusion part, other convolution operations are all identical to the DWParNet shown in Fig.~\ref{fig_dw}.

\begin{table}[tbp]
\renewcommand{\arraystretch}{1.3}  
\caption{The hyperparameters of the proposed model}
\label{table_hyper_model}
\centering
\resizebox{0.3\textwidth}{!}  
{
\begin{tabular}{ll}
\toprule
\textbf{Hyperparameter} & \textbf{Values}\\
\midrule
Branch1-Conv1 & 64\\
Branch2-Conv1 & 128\\
Branch2-Conv2 & 256\\
Branch3-Conv1 & 32\\
Branch3-Conv2 & 96\\
Branch3-Conv3 & 192\\
Fusion-Conv & 192\\
LSTM-Layer & 2\\
LSTM-Hidden Size & 32\\
Fusion-Pooling & Adaptive avg\\
Fusion-Dropout & 0.5\\
\bottomrule
\end{tabular}
}
\end{table}

\subsection{Resource Adaptation Based on Vehicular Environment}
\label{allocation model}
Through different down-sampling processing of sub-neural networks, the number of parameters to be processed and the memory usage are different for each sub-neural network. The less computation of the sub-network, the fewer ECU resources and less running time it occupies. Similarly, different ECU devices in the original vehicle network will produce different resource occupancy according to the functional modules in charge. 

To make full use of numerous ECU resources in the vehicle, we also propose a resource adaptation algorithm based on the in-vehicle CAN bus. The algorithm mainly quantifies the importance and resources of each ECU module and collects indicators from three aspects: processor idle rate, memory idle rate, and importance level of functional modules in ECUs. Then, we analyze the memory occupancy and calculation rate of each branch network on the corresponding ECU, and define the resource occupancy index of the branch network. We compare the branch network occupancy index with the ECU availability index to determine whether to allocate the branch network to the corresponding ECU. The mathematical model we constructed is as follows.

In this method, $n$ shallow sub-neural network structures are allocated to $n$ different ECUs to run. Let the total number of ECUs in the vehicle be $N$, and the processor idle rate and the memory idle rate of the $i$ th ECU be $P_{i}$ and $M_{i}$ when it is running normally. Then, the total resource idle rate $S_{i}$ of the ECU can be comprehensively evaluated and defined as follows:

\begin{equation}
S_{i} =\frac{2P_{i} \cdot M_{i}}{P_{i} + M_{i}}.
\label{eq_S} 
\end{equation}

Automobile manufacturers always have a risk rating for the car's module system, indicating the risk level the module malfunction could lead to. We use these risk indexes to describe the importance index of the module, set as a positive integer $R_{i}$ (usually no more than 10). The higher the index shows the higher the importance index is, and the more inappropriate it is to be assigned the task of a branch network. Therefore, the overall availability index $U_{i}$ of the $i$ th ECU is defined as follows:

\begin{equation}
U_{i} =\left ( 1+\alpha^{2} \right ) \frac{S_{i} \cdot \frac{1}{R_{i}} }{\alpha^{2} \cdot \frac{1}{R_{i}}+ S_{i}},
\label{eq_U} 
\end{equation}
where $\alpha$ is a positive integer coefficient, used to adjust the balance between the importance and the resource idle rate. Then, by substituting Equation (\ref{eq_S}) into Equation (\ref{eq_U}) and simplify, we have:

\begin{equation}
U_{i} =\frac{2\left ( 1+\alpha^{2} \right ) \cdot P_{i}\cdot M_{i} }{\alpha^{2}\cdot \left ( P_{i}+ M_{i} \right )+2P_{i}\cdot M_{i}\cdot R_{i} }.
\label{eq_U2} 
\end{equation}

In addition, we define the memory occupation ratio of the $j$ th branch network to be allocated on the $i$ th ECU as $m_{ij} \left(m_{ij} < 1\right)$, that is, the ratio of the memory occupied by the $j$ th branch network to the total memory of the $i$ th ECU. Since the main calculation amount of the learning network model is forward/backward propagation, the calculation amount of the forward/backward propagation process model can reflect the situation of CPU resource occupation. We define the calculation rate of the $j$ th branch network structure as:

\begin{equation}
c_{j}=\frac{Forward/backward \ pass \ size}{Total \ model \ size}.
\label{eq_c}
\end{equation}

Based on a comprehensive evaluation, the resource occupation index of the $j$ th branch network on the $i$ th ECU is defined as:

\begin{equation}
O_{ij}=\frac{\left ( 1+\beta ^2 \right )\cdot c_{j} \cdot m_{ij}}{\beta^{2} \cdot m_{ij} + c_{j}} ,
\label{eq_O}
\end{equation}
where $\beta$ is a positive integer coefficient, used to adjust the balance between the memory occupation ratio and the calculation rate.

The algorithm compares $U_{i}$ with $O_{ij}$ to determine whether the task of the $j$ th branch network is suitable to be assigned to the $i$ th ECU. If $U_{i} \ge O_{ij}$, the $i$ th ECU can be selected to install the $j$ th branch shallow neural network, otherwise the branch module cannot be loaded onto the ECU.

\section{Experimental results and performance evaluation}
\label{Section4}
\subsection{Experimental setup}
This experiment was carried out on a MacBook Pro notebook. The hardware conditions include a 2.2GHz quad-core Inter Core i7 processor, 16GB 1600MHz DDR3 memory, and an Inter Iris Pro 1536MB graphics card. The experiment is based on the operating system of MacOS Monterey version 12.6 and PyCharm 2022.2.2 Community version, in which Python 3.9.12 and Pytorch 1.12.1 are installed.

\subsection{Data pre-processing}
The dataset we used is the Car-Hacking dataset \cite{GIDS}, which is obtained by recording CAN traffic through OBD-II port in a real vehicle and includes one normal dataset and four different attack datasets: DoS, Fuzzy, spoofing GEAR and spoofing RPM. Since the model proposed in this study combines branches of two types of network structure, respectively for spatial feature extraction and temporal feature extraction, there are two forms of data for branch network input.

First, the in-vehicle CAN message data is processed into image form. We extract CAN ID and 8-byte payload from CAN message data to form a feature vector of a CAN message with 9 feature values. Then, every 9 feature vectors are taken to form a 9 * 9 two-dimensional feature matrix which is taken as a channel of an image, and every 3 channels form an image which is stored in a three-dimensional tensor. That is, the shape of each input data for the convolution branch is 3 * 9 * 9. The image data distribution after pre-processing is shown in Table~\ref{table_data_partition}. If the image is entirely composed of normal messages, it will be marked as normal data, otherwise, it will be marked as the corresponding type of attack data.

For the temporal feature extraction branch, the three-dimensional tensor will be expanded and reshaped into a two-dimensional tensor with a shape of 27 * 9. Every 27 vectors are input into the LSTM network as a sequence with a temporal relationship. It should be noted that the LSTM network structure needs to learn the temporal characteristics between CAN messages, so the sequence of 27 data in each group that form an image cannot be disturbed during data processing.

\begin{table}[tbp]
\renewcommand{\arraystretch}{1.3}
\caption{The dataset partition for training, validation, and testing}
\label{table_data_partition}
\centering
\resizebox{0.48\textwidth}{!}
{
\begin{tabular}{llll}
\toprule
\textbf{Type} & \textbf{Training set} & \textbf{Validation Set} & \textbf{Testing set}\\
\midrule
Normal & 25,637 & 7,325 & 3,662\\
DoS & 28,145 & 8,042 & 4,020\\
Fuzzy & 33,439 & 9,554 & 4,776\\
Spoofing Gear & 49,063 & 14,018 & 7,009\\
Spoofing RPM & 53,646 & 15,328 & 7,663\\
\bottomrule
\end{tabular}
}
\end{table}

\subsection{Model training}
\begin{figure}[tbp]
\centering
\subfigure[The training loss and validation loss]{
\label{fig_loss}
\includegraphics[width=2.5 in]{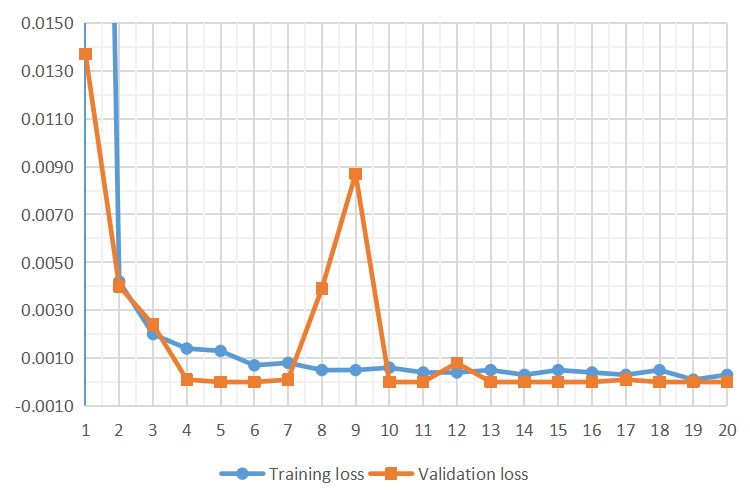}}  
\subfigure[The training accuracy and validation accuracy]{
\label{fig_accuracy}
\includegraphics[width=2.5 in]{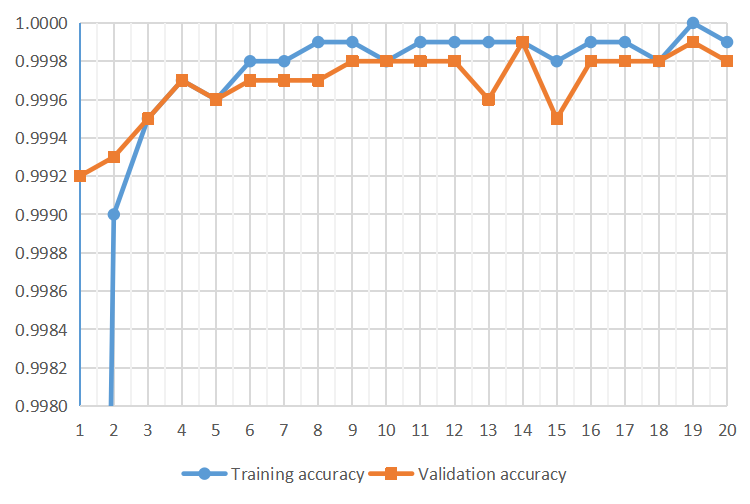}}  
\caption{The curves of loss and accuracy during model training.}
\label{fig_training}
\end{figure}

We have adopted a validation dataset to determine the best hyperparameters of the model and model training. The trend of loss value and accurate value during cross-validation is shown in Fig.~\ref{fig_training}. According to the changing trend of loss and accuracy, we decide to take 14 as the best training epoch because there is not too much performance improvement of the model after epoch 14 and further training will lead to over-fitting. We train the proposed STParNet on the Adam optimizer with a 0.0001 initial learning rate and use a sparse categorical cross-entropy to calculate the loss value. All the best-performing hyperparameters for model training are summarized in Table~\ref{table_hyper_train}. 

\begin{table}[tbp]
\renewcommand{\arraystretch}{1.3}  
\caption{The hyperparameters for training the model}
\label{table_hyper_train}
\centering
\resizebox{0.5\textwidth}{!}  
{
\begin{tabular}{ll}
\toprule
\textbf{Hyperparameter for training} & \textbf{Values}\\
\midrule
Learning rate & 0.0001\\
Optimizer & Adam\\
Loss function & sparse categorical cross-entropy\\
Batch size & 32\\
Epoch & 14\\

\bottomrule
\end{tabular}
}
\end{table}

\subsection{Evaluation metrics}
In this study, we not only evaluate the detection ability and generalization ability of the model but also evaluated the lightweight degree and running efficiency of the model. We used the confusion matrices, top-1 accuracy, and AUC values to evaluate the detection effect of models. The AUC is selected because it can reflect the processing ability of the model for the unbalanced dataset, that is, the generalization ability of the model. The calculation method of top-1 accuracy is

\begin{equation}
Accuracy = \frac{1}{N}\sum_{i}^{N}1\left ( y_{i} = \hat{y}_{i} \right ) .
\label{eq_accuracy}
\end{equation}

To evaluate the lightweight degree of the model, we used the total parameter quantity and total memory consumption to describe the model size. Among them, the total memory consumption is about the sum of forward/backward pass usage and parameter usage. In addition, training speed and inference speed are used to evaluate the running efficiency of the model.

\subsection{Results and analysis}
\begin{figure}[tbp]
\centering
\subfigure[The confusion matrix of DWParNet]{
\label{fig_conf_DW}
\includegraphics[width=3.4in]{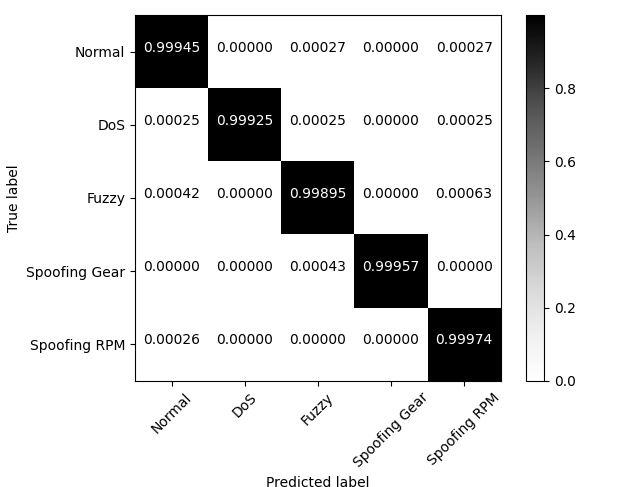}}  
\subfigure[The confusion matrix of STParNet]{
\label{fig_conf_ST}
\includegraphics[width=3.4in]{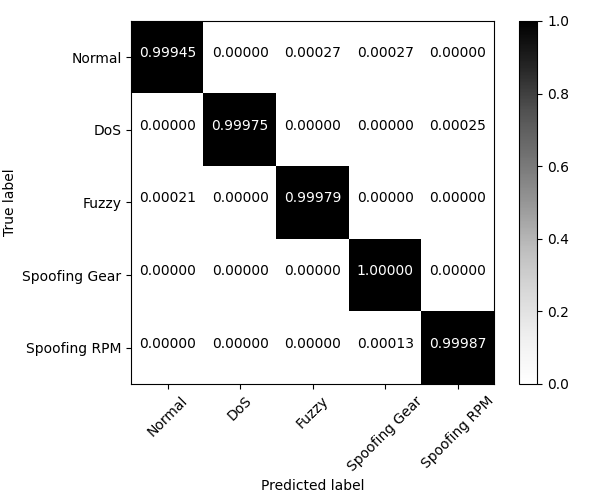}}  
\caption{The detection effect of DWParNet and STParNet.}
\label{fig_conf}
\end{figure}

One of the purposes of our research is to verify whether the integration of temporal and spatial feature learning is better than the single spatial feature learning structure. Therefore, we compare and analyze the intrusion detection results of DWParNet and STParNet. The results are shown in Fig.~\ref{fig_conf}. Both DWParNet and STParNet have great detection accuracy of each kind of attack data and normal data, all higher than 0.9998. Compared with the single spatial feature learning network DWParNet, the STParNet proposed in this paper has improved the detection and classification performance of attacks. The classification accuracy of the four attacks has been improved by 0.00050, 0.00084, 0.00043 and 0.00013 respectively, which illustrates that 
adding LSTM structure to extract temporal features and analyzing with spatial features can effectively improved the detection ability of the model compared with only learning data spatial features.

\begin{figure}[tbp]
\centering
\includegraphics[width=3.4in]{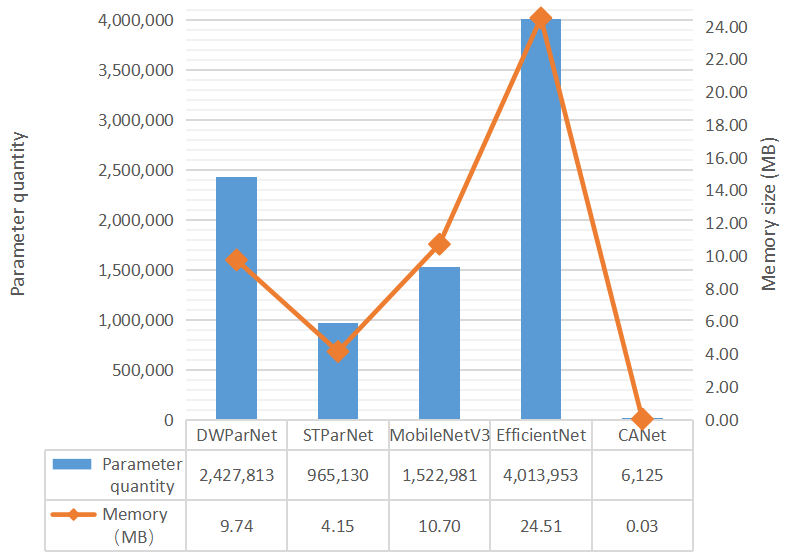} 
\caption{The parameter quantity and memory consumption of the models.}
\label{fig_lightweight}
\end{figure}

\begin{figure}[t]
\centering
\includegraphics[width=3.4in]{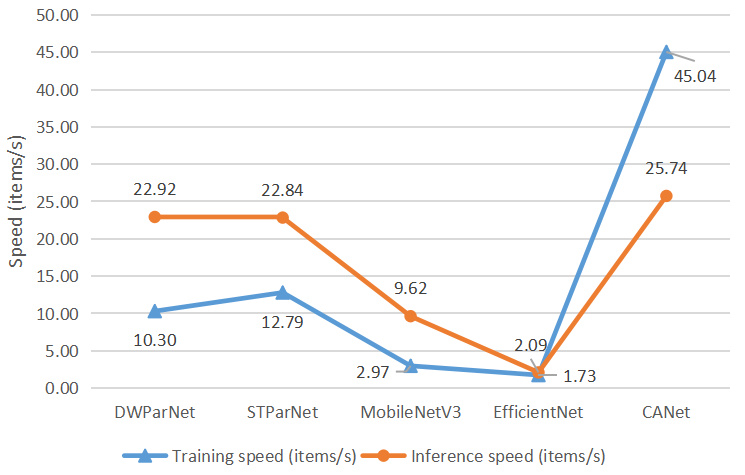} 
\caption{The training speed and inference speed of the models. ``item'' refer to the data that the model can process at one time and ``items/s'' refers to the batches of data that the model can process per second.}
\label{fig_running}
\end{figure}

Moreover, we select three baseline methods for comparative experiments: MobileNetV3 \cite{MobileNetV3}, EfficientNet \cite{EfficientNet}, and CANet \cite{CANet-LSTM-autoencoder}. Although MobileNetV3 and EfficientNet are not models developed for vehicular environments, nor are they models for intrusion detection, they are the most classic models of deep learning algorithms developed from cloud to end, and from deep learning to lightweight learning. They all have advanced performance and small model sizes and are the best among the lightweight models of deep learning networks so far. CANet is the most lightweight model with better performance in the field of in-vehicle CAN bus intrusion detection, which is selected by a large number of experiments in the research \cite{our-analysis}. The comparative experimental results of detection performance, lightweight degree, and running efficiency are shown in Table~\ref{table_detection_res}, Fig.~\ref{fig_lightweight}, and Fig.~\ref{fig_running}, respectively.

\begin{table}[tbp]
\renewcommand{\arraystretch}{1.3}  
\caption{The results of detection performance}
\label{table_detection_res}
\centering
\resizebox{0.35\textwidth}{!}  
{
\begin{tabular}{lll}
\toprule
\textbf{Models} & \textbf{Accuracy}  & \textbf{AUC}\\
\midrule
DWParNet (Ours) & 0.9994 & 0.99981\\
STParNet (Ours) & \textbf{0.9998} & \textbf{1.00000}\\
MobileNetV3& 0.9990 & 0.99990\\
EfficientNet & 0.9993 & 0.99996\\
CANet & 0.9942 & 0.92146\\

\bottomrule
\end{tabular}
}
\end{table}

These experimental results show that the STParNet proposed in our study has the highest detection accuracy and AUC value, meaning that it has the greatest intrusion detection performance among these models. Compared with MobileNetV3 and EfficientNet, it can be seen that STParNet can greatly reduce the model parameter quantity and memory usage to half and a quarter of the two models' original sizes respectively, which has been able to meet the constraints of in-vehicle resources. Also, STParNet has obvious advantages in both training speed and inference speed, about two times of MobileNetV3's speed and ten times of EfficientNet's speed. For DWParNet, the training speed of STParNet is also slightly improved, from 10.30 to 12.79 items/s. Compared with the CANet model, STParNet still has a gap in the size and running speed of the model, but the detection effect is much better than CANet, improving 0.0056 in accuracy, and especially, its generalization ability described by AUC value is improved from about 0.92146 to 1.00000, which means that it can provide better security for the in-vehicle network.

Furthermore, we calculate and analyze the resource consumption of each branch network of the STParNet. The parameter quantity and calculation quantity of each branch network are collected through experiments, the results are shown in Table~\ref{table_branch}. The four branch structures will be allocated to four different ECUs according to the algorithm. The fusion part is undertaken by the central electronic module. That is because the task load of fusion is relatively large and the calculation is concentrated, and the central electronic module has more resources than ECU and can fully undertake the feature fusion part. The results prove that the design of parallel architecture can greatly reduce the model size and resource consumption for each ECU, even less than 0.5MB, while ensuring detection capability. Among them, the size and task load of branch 4 are 0.06MB which is similar with that of the CANet model, about 0.03MB, but from the perspective of the intrusion detection effect, STParNet has more advantages and higher security as mentioned before.

According to the mathematical model proposed in Section~\ref{allocation model}, the calculation rates of the four branch networks are 0.8889, 0.6216, 0.5833, and 0.1667, calculated by Equation (\ref{eq_c}). We assumed that the total size of each ECU memory is 1MB. Then, calculated by Equation (\ref{eq_O}), their memory occupancy rates are 0.09, 0.37, 0.24, and 0.06. Based on experience, the memory occupation ratio has a greater impact on the resource occupation index, so we set $\beta = 2$. After calculation, the resource occupation indexes of the four branches are 0.3203, 0.5472, 0.4535, and 0.1230. The result illustrates that the four branches all have a very low demand for available resources of ECU. As long as $U_{i} \ge O_{ij}$ is satisfied, these four branches can be allocated to the corresponding ECU.
\newcommand{\tabincell}[2]{\begin{tabular}{@{}#1@{}}#2\end{tabular}}

\begin{table*}[tbp]
\renewcommand{\arraystretch}{1.3}  
\caption{The statistical results of each branch network of STParNet and baseline CANet}
\label{table_branch}
\centering
\resizebox{1\textwidth}{!}  
{
\begin{tabular}{llllll}
\toprule
\textbf{Branch/model} & \textbf{Forward/backward pass size (MB)} 
& \textbf{Parameter size (MB)} & \textbf{Total size (MB)} 
& \textbf{Calculation rate $c_j$} & \textbf{Occupation index $O_{ij}$}\\
\midrule
Branch1-Conv & 0.08 & 0.00 & 0.09 & 0.8889 & 0.3203\\
Branch2-Conv & 0.23 & 0.15 & 0.37 & 0.6216 & 0.5472\\
Branch3-Conv & 0.14 & 0.10 & 0.24 & 0.5833 & 0.4535\\
Branch4-LSTM & 0.01 & 0.05 & 0.06 & 0.1667 & 0.1230\\
Fusion & 0.01 & 3.38 & 3.40 & / & /\\
STParNet (Ours) & 0.47 & 3.68 & 4.15 & / & /\\
CANet & 0.01 & 0.02 & 0.03 & 0.3333 & 0.1103\\
MobileNetV3 & 4.20 & 5.81 & 10.70 & / & /\\
EfficientNet & 8.52 & 15.31 & 24.51 & / & /\\

\bottomrule
\end{tabular}
}
\end{table*}

\section{Related works}
\label{Section5}
The IDSs for the in-vehicle CAN bus can be divided into four types: (1) the specification-based approaches which detect abnormal behavior that does not match system specifications, such as protocols and frame formats \cite{specification-based}; (2) the fingerprint-based approaches which use contours defined based on ECU features rather than human-defined specifications, such as the clock skew of ECUs \cite{clock-skew-ECU} and the voltage fingerprint \cite{voltage-based1,voltage-based}; (3) the statistics-based approaches, such as entropy-based \cite{entropy-based} and frequency-based \cite{frequency-based}; (4) the learning-based approaches, such as \cite{CANintelliIDS, VehCom-VehicularIDS-DCNN}, and \cite{SAC-VehicularIDS-CANTransfer}, which perform better than other types without prior expert knowledge, especially deep learning-based methods, but rely on sufficient computing resources.

In order to enable the IDSs to be installed on the terminal so that it can achieve more efficient and stable detection capability, the existing research direction of learning-based technology has been developing towards lightweight models. There are two kinds of lightweight directions we have investigated in the field of IDS: (1) using a simple and shallow CNN structure combined with other technologies to enhance the detection performance, such as combining an LSTM structure \cite{HyDL-IDS-CNN-LSTM} and using recursive graphs to process the data \cite{Rec-CNN}, and (2) using autoencoder structure with lightweight neural units, such as \cite{CANet-LSTM-autoencoder, WCL-LDAN}, and \cite{TIA-VehicularAutoEncoder}. In the research \cite{our-analysis}, these models have been analyzed in the same experimental environment, and it is found that CANet has the best comprehensive performance. Therefore, we chose the CANet as one of the baseline methods in our experiment.

In fact, the development direction of deep learning model lightweight is not only in the field of IDS but also in the field of deep learning image processing which began the research on lightweight structure earlier. Two of the lightweight technologies are to use the DW convolution instead of traditional convolution and to adopt Squeeze-and-Excitation (SE) modular structure. They are both mentioned in \cite{MobileNetV3} and \cite{EfficientNet}.
In addition, channel shuffle and channel split can also be used to improve the running efficiency of deep learning models, such as \cite{shuffleNetV2}. 

For parallel network structure, although there is no corresponding application in the field of IDS, there are parallel structures and models of neural network learning in the field of image processing. For example, the Inception structure proposed in research \cite{GoogleNet} is a parallel structure, although the whole network is still composed of these Inception structures that are deeply stacked. Or, the algorithm parallelizes the learning process of a deep network, such as \cite{par-machine-learning}. The one using parallel branches as network structure we found is the ParNet proposed by Ankit et al. in \cite{ParNet}.

\section{Conclusion}\label{Section6}
We have proposed a lightweight intrusion detection model, LiPar, for the in-vehicle CAN bus, including the STParNet and the resource adaptation algorithm. Through comparative experiments on STParNet and DWParNet, we have proved that the fusion of temporal and spatial feature extraction structures can effectively improve the intrusion detection accuracy and the generalization ability of the model. Moreover, we found that the design of parallel structures with multi-dimension feature extraction can help to enhance a lot of the detection performance as well as reduce the model size and task load for each ECU. We have also proposed a resource adaptation algorithm to allocate the branch task to the appropriate ECU. After calculating, the resource occupancy indexes also illustrate that the branched STParNet are lightweight enough to easily find ECU suitable for allocation. In conclusion, the LiPar model we proposed has great advantages in terms of detection performance, generalization ability, running efficiency, and lightweight degree. It can be loaded on the vehicle practically and be a better choice for the in-vehicle CAN bus intrusion detection system.

\section*{Acknowledgments}
This work is supported by National Natural Science Foundation of China (NSFC) (grant number 62272129) and Taishan Scholar Foundation of Shandong Province (grant number tsqn202408112).


\vspace{-22pt}
\begin{IEEEbiography}[{\includegraphics[width=1in,height=1.25in,clip,keepaspectratio]{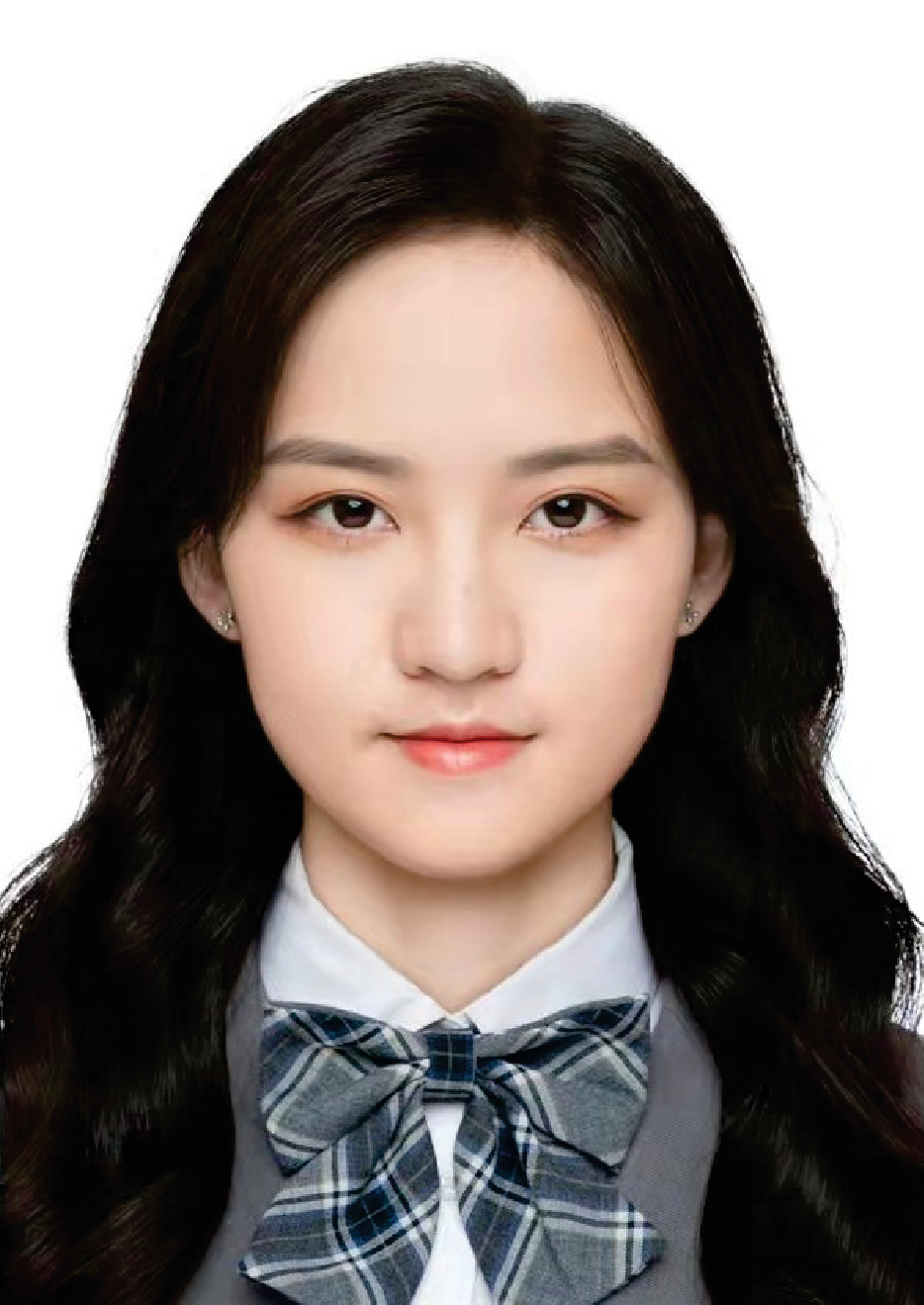}}]{Aiheng Zhang}
received the B.S. degree in computer science and technology from the Beijing University of Technology, Beijing, China. She is currently pursuing a master’s degree in computer technology with the Harbin Institute of Technology (HIT), China. Her research interests include intelligent and lightweight in-vehicle intrusion detection models.
\end{IEEEbiography}

\vspace{-22pt}
 
\begin{IEEEbiography}[{\includegraphics[width=1in,height=1.25in,clip,keepaspectratio]{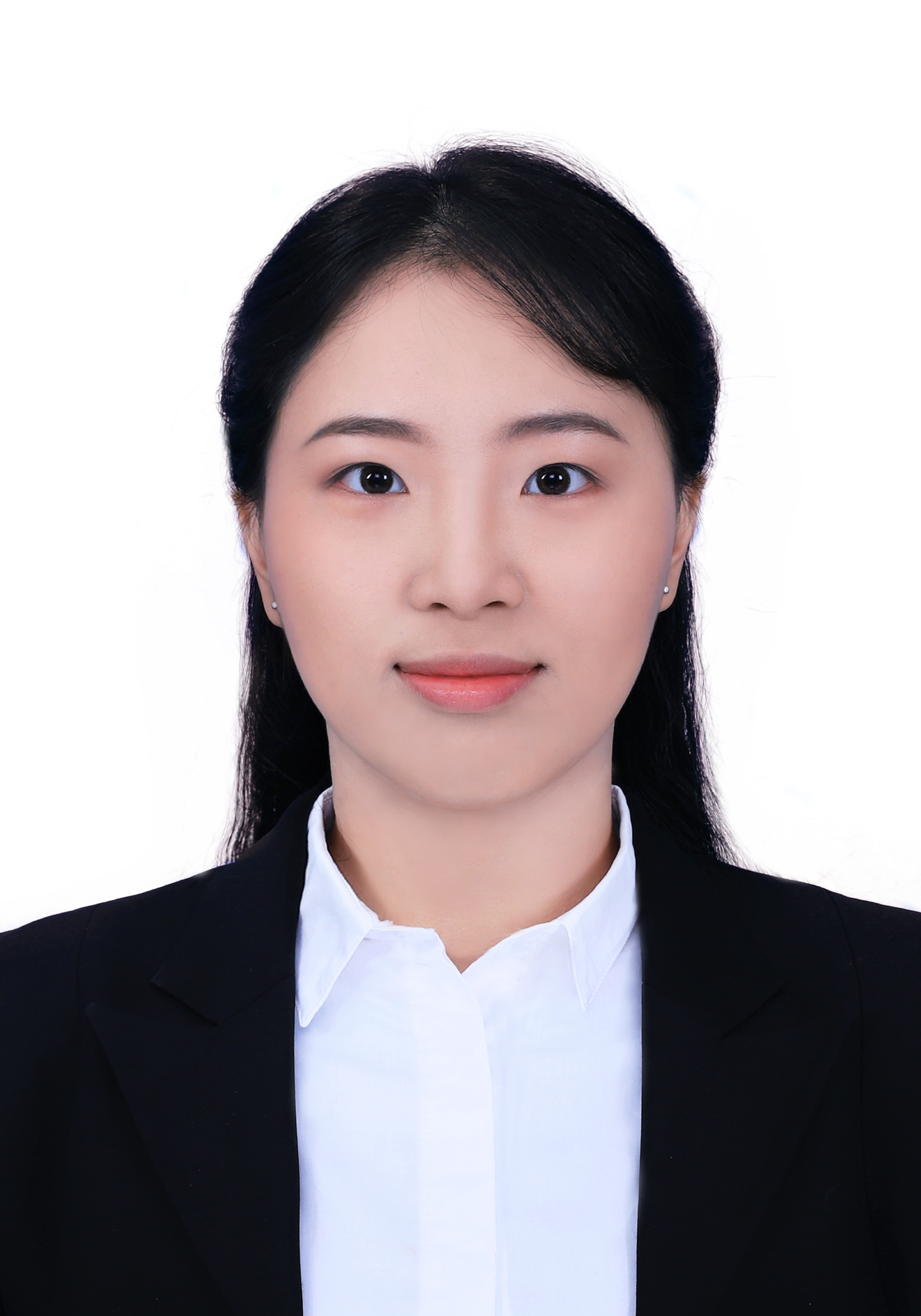}}]{Qiguang Jiang}
received the B.S. degree in Information and Computing Science from the Harbin Institute of Technology (HIT), Weihai, China. She is currently pursuing the master's degree in computer science and technology with the Harbin Institute of Technology (HIT), China. Her research interests include intelligent and efficient in-vehicle intrusion detection models.
\end{IEEEbiography}

\vspace{-22pt}
\begin{IEEEbiography}[{\includegraphics[width=1in,height=1.25in,clip,keepaspectratio]{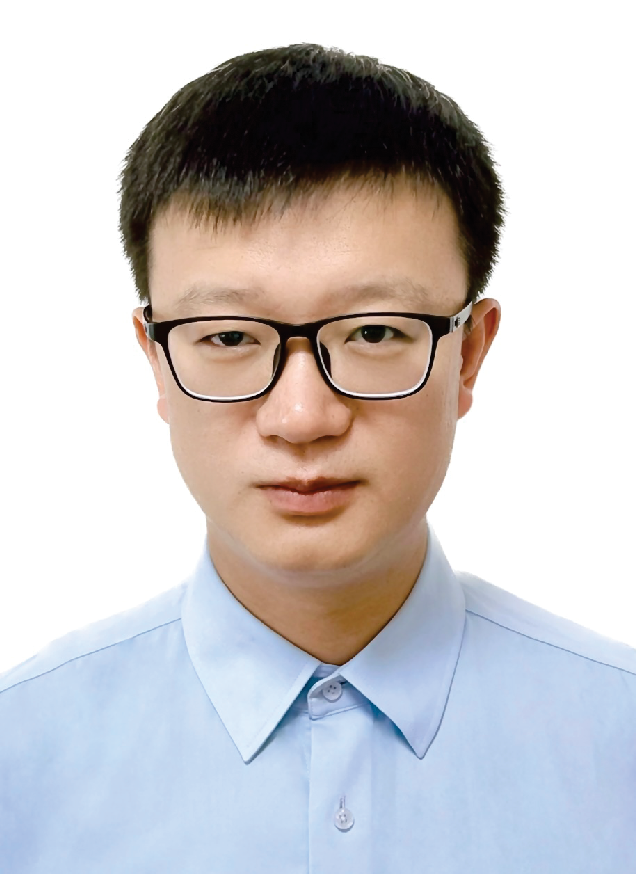}}]{Kai Wang}
received the B.S. and Ph.D. degrees from Beijing Jiaotong University. He is currently a Professor with the School of Computer Science and Technology, Harbin Institute of Technology (HIT), Weihai. Before joined HIT, he was a postdoc researcher in computer science and technology with Tsinghua University. He has published more than 40 papers in prestigious international journals, including IEEE TITS, IEEE TCE, ACM TOIT, ACM TIST, etc. His current research interest is on security for cyber-physical systems and emerging networks (e.g., autonomous vehicles, industrial control systems, IoT), and applied machine learning for network attacks detection and information forensics. He is a Member of the IEEE and ACM, and a Senior Member of the China Computer Federation (CCF).
\end{IEEEbiography}
\vspace{-22pt}
\begin{IEEEbiography}[{\includegraphics[width=1in,height=1.25in,clip,keepaspectratio]{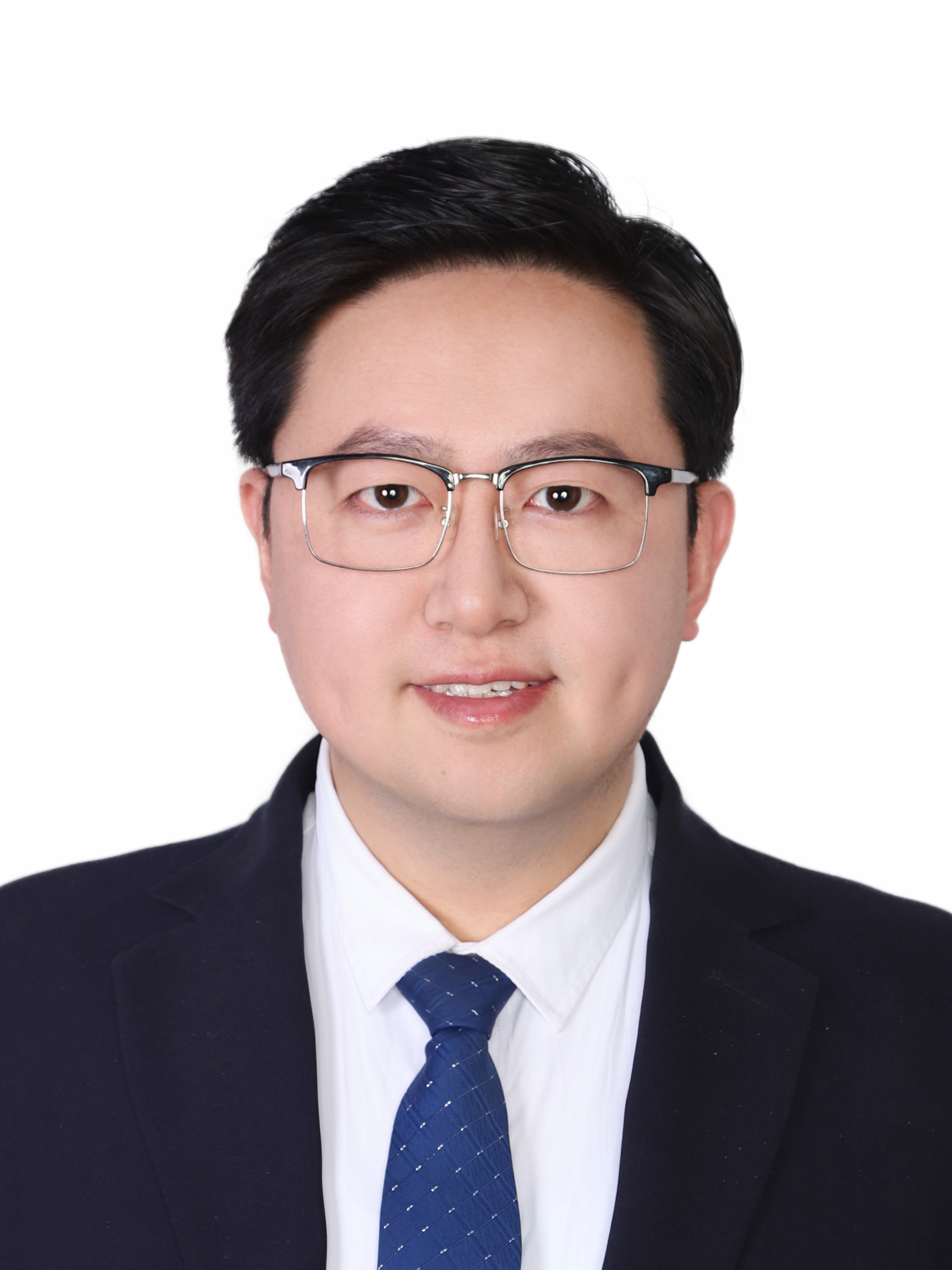}}]{Ming Li}
obtained his B.S. degree from Shandong University, M.Sc. degree from Ulm University, and Ph.D. degree from Hamburg University of Technology. He is currently the Vice General Manager and CTO at Shandong Inspur Database Technology Co., Ltd. Before joining the Inspur Group, he worked as a senior engineer at Intel Corporation. He has published over 10 papers and books, including IEEE GLOBOCOM, ICC, ITC, WCNC, VTC, etc., and has also participated in several national standards. His current research interests include HTAP cloud-native high-performance databases, big data analytics systems, and edge computing. He serves as an Executive Director of the China Industry-University-Research Institute Collaboration Association (CIUR) and is the Director of the Jinan Key Laboratory of Distributed Databases.
\end{IEEEbiography}

\end{document}